\documentclass[12pt]{iopart}

\usepackage{iopams}  
\usepackage[dvipdfmx]{xcolor}
\usepackage[pdftex]{graphicx} 

\begin{document}

\title[Peculiarity of Symmetric Ring Systems with Double Y-Junctions]
{Peculiarity of Symmetric Ring Systems with Double Y-Junctions and the 
magnetic effects}

\author{Yukihiro Fujimoto$^1$, Kohkichi Konno$^{2, 3}$, Tomoaki Nagasawa$^2$}

\address{$^1$ National Institute of Technology, 
 Oita College, 1666 Maki, Oita 870-0152, Japan}
\address{$^2$ National Institute of Technology, 
 Tomakomai College, 443 Nishikioka, Tomakomai 059-1275, Japan}
\address{$^3$ Department of Applied Physics, 
 Hokkaido University, Sapporo 060-8628, Japan}

\ead{y-fujimoto@oita-ct.ac.jp (Yukihiro Fujimoto), kohkichi@tomakomai-ct.ac.jp (Kohkichi Konno), 
nagasawa@tomakomai-ct.ac.jp (Tomoaki Nagasawa)}
\begin{indented}
\item[]
\end{indented}

\begin{abstract}

We discuss quantum dynamics in the ring systems 
with double Y-junctions in which two arms have same length. 
The node of a Y-junction can be parametrized by U(3).
Considering mathematically permitted junction conditions seriously,  
we formulate such systems by scattering matrices.
We show that the symmetric ring systems, which consist 
of two nodes with the same parameters under the reflection symmetry, 
have remarkable aspects that there exist localized states 
inevitably, and resonant perfect transmission occurs 
when the wavenumber of an incoming wave coincides 
with that of the localized states, for any parameters 
of the nodes except for the extremal cases in which 
the absolute values of components of scattering matrices take $1$.
We also investigate the magnetic disturbance to the 
symmetric ring systems.
\end{abstract}

%
%
\submitto{\JPA}
%
%
%

\section{Introduction}

Quantum mechanical features in non-simply connected, 
ring geometries have attracted great interest from many physicists.  
Particularly striking phenomena in ring systems 
are induced by the Aharonov-Bohm effect \cite{ab} 
and the Aharonov-Casher effect \cite{ac}.
To prove these effects, ring systems have been realized
in semiconductor nanotechnology (see e.g., \cite{webb_etal,tonomura_etal}).
More recently, various physical characteristics
of quantum ring systems have also been investigated in
\cite{fuhrer_etal,vksm,ihn_etal}. 
Furthermore, applications of ring systems 
to a qubit were discussed \cite{rasanen_etal, zksm}.
A typical structure of ring systems is formed by 
connected double Y-junctions. A Y-junction is composed of three 
one-dimensional quantum wires intersecting at one point (i.e. a node).
We shed light on such quantum ring systems with 
double Y-junctions.

Quantum ring systems with Y-junctions were originally 
investigated in the pioneer theoretical works \cite{bia,buttiker}.
At a node, a simple form of the scattering matrices 
had been customary assumed for years in the literature.
On the other hand, mathematical features of point interactions 
at a node were thoroughly investigated in \cite{rs,seba,aghh,cft}.
These works showed that the point interaction on a
Y-junction can be parametrized by U(3).
In subsequent works, certain aspects of transmission properties of 
a quantum particle in the system of a single Y-junction 
were studied in \cite{cet},  
and a star graph and related topics were also discussed 
in \cite{tm,aj,ohya,tc}.  Ring systems with double Y-junctions
were restudied in \cite{fknt} based on the mathematical framework.
However, the previous work \cite{fknt}
is restricted to a subclass in parameter space of U(3), 
i.e., the scale invariant class.
Therefore, in this paper, we provide a more general discussion 
of quantum ring systems with double Y-junctions 
without the tight restriction on the parameter space.

We formulate quantum dynamics in the ring systems with double Y-junctions,
taking account of mathematically permitted junction conditions seriously. 
While we assume that the two arms of a ring has the same length, 
we do not impose any restriction on the parameters of two nodes 
from the beginning. From our analysis, we find that the symmetric ring
systems, which have two same nodes under the reflection symmetry, 
have remarkable features that localized states exist inevitably
and that resonant perfect transmission 
occurs for any parameters of the nodes except for the extremal cases 
in which the absolute values of components of scattering matrices 
take $1$. 
We also discuss disturbance to the symmetric ring systems.
In particular, we investigate the effect of magnetic flux which penetrates 
the ring systems. 
From our analysis, we find the general expressions of the amplitudes 
for reflection and transmission in the presence of the magnetic flux.
This paper is organized as follows. In Sec.~\ref{sec:formulations}, 
we review the formulation by scattering matrices based on \cite{fknt}.
In Sec.~\ref{sec: l_states}, we investigate localized states 
and show that the existence of the localized states is 
inevitable in the symmetric ring systems.
In Sec.~\ref{sec:trans}, focusing on scattering problems in the symmetric ring systems,
we investigate the transmission probability through the ring systems.
Then we find that the resonant perfect transmission occurs 
when the wavenumber of an incoming wave coincides 
with that of localized states. 
In Sec.~\ref{sec:md}, we consider external magnetic fields
as disturbance to the ring systems. 
Formulating the quantum ring systems in the presence
of the magnetic fields, we investigate probability amplitudes
for reflection and transmission.
Finally, we give a conclusion in Sec.~\ref{sec:conclusion}.

\section{Formulation of a quantum particle
in a ring system with double Y-junctions}
\label{sec:formulations}

\subsection{The coordinate system and the basic equation}

\begin{center}
\begin{figure}[h]
  \includegraphics[width=0.8\textwidth]{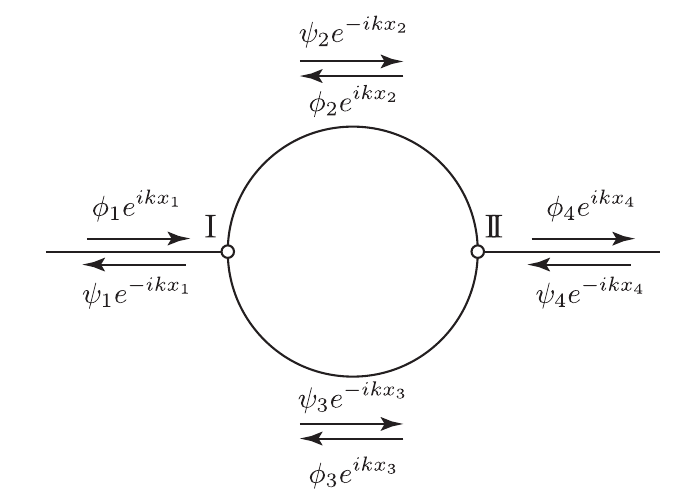}
  \caption{\label{fig1} 
  A ring system with double Y-junctions. This ring has two nodes (I and I$\!$I).
  Plane waves on this system are expressed by 
  $\phi_{1} e^{ikx_{1}}$, $\psi_{1} e^{-ikx_{1}}$, 
  $\phi_{2} e^{ikx_{2}}$, $\psi_{2} e^{-ikx_{2}}$, 
  $\phi_{3} e^{ikx_{3}}$, $\psi_{3} e^{-ikx_{3}}$, 
  $\phi_{4} e^{ikx_{4}}$, and $\psi_{4} e^{-ikx_{4}}$.}
\end{figure}
\end{center}

\begin{center}
\begin{figure}[h]
  \includegraphics[width=.8 \textwidth]{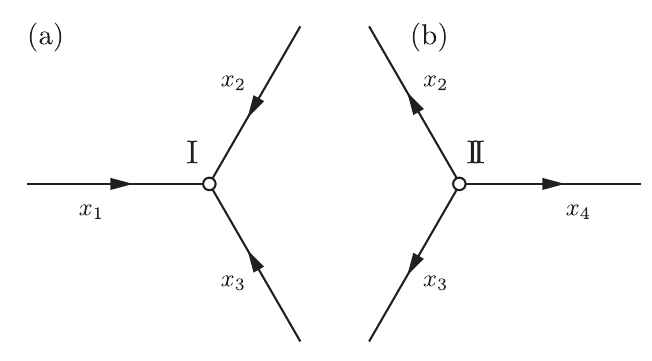}
  \caption{\label{fig2} (a) The coordinate axes for the node I.
  (b) The  coordinate axes for the node I$\!$I.}
\end{figure}
\end{center}

We discuss a ring system with double Y-junctions as shown in Fig.~\ref{fig1}.
Three one-dimensional quantum wires intersect at one point (i.e., node)
in each Y-junction. We describe the Y-junction on the 
left-hand side by the inward coordinate axes $x_{1}, x_{2}$ and $x_{3}$,  
and that on the right-hand side by the outward coordinate axes 
$x_{2}, x_{3}$ and $x_{4}$,  as shown in Fig.~\ref{fig2}(a) and (b), respectively.
Note that the angle between any two axes and the curvature 
of the wires have no effect on the physical states.
We assume that the nodes locate at $x_{i}=\xi_{\rm \rm I}$ ($i=1, 2, 3$) 
and at $x_{j}=\xi_{\rm \rm I\!I}$ ($j=2, 3, 4$), 
where $\xi_{\rm \rm I} > \xi_{\rm \rm I\!I}$. 
We consider a free quantum particle on this system, 
which obeys the Schr\"odinger equation on the wires,
\begin{equation}
\label{eq:se}
 i\hbar \frac{\partial }{\partial t} \Phi_{i} (t, x_{i})
 = - \frac{\hbar^2}{2m} \frac{\partial^2}{\partial x_{i}^2} \Phi_{i} (t, x_{i}) 
 \quad (i= 1, 2, 3, 4),
\end{equation}
where $\Phi_{i}$ denotes the wave function on the $x_{i}$-axis, and 
$m$ denotes the mass of the particle.

\subsection{Junction conditions}

At the nodes, we have to impose the junction conditions,
which are provided by the conservation of
probability current, i.e., 
\begin{eqnarray}
\label{eq:jc1}
 & \displaystyle \sum_{i=1}^{3}  
 \lim_{x_{i} \rightarrow \xi_{\rm I}^-}  j_{i} (t, x_{i} ) = 0 , 
 \\ 
\label{eq:jc2}
 & \displaystyle \sum_{i=2}^{4}  
 \lim_{x_{i} \rightarrow \xi_{\rm I\!I}^+}  j_{i} (t, x_{i} ) = 0 ,
\end{eqnarray}
where $x_{i} \rightarrow \xi_{\rm I}^-$ denotes the limit from below, and
$x_{i} \rightarrow \xi_{\rm I\!I}^+$ denotes the limit from above.
The probability current $j_{i}$ is defined by
\begin{equation}
 j_{i} (t, x_{i} ) := -\frac{i\hbar}{2m} 
 \left\{ \Phi^{\ast}_{i} (t, x_{i})  \partial_{x} \Phi_{i} (t, x_{i}) 
         - \left(  \partial_{x} \Phi^{\ast}_{i} (t, x_{i})  \right) \Phi_{i} (t, x_{i}) \right\} .
\end{equation}
The junction conditions (\ref{eq:jc1}) and (\ref{eq:jc2}) 
can be expressed by \cite{cft}
\begin{equation}
\label{eq:jc3}
 \Psi'^{\dagger} \Psi - \Psi^{\dagger} \Psi'=0 ,
\end{equation}
where for the Y-junction $(x_{1}x_{2}x_{3})$ (node I), we should take 
\begin{equation}
\label{eq:def-Psi1}
 \Psi := \lim_{x\rightarrow \xi_{\rm I}^-} \left(
 \begin{array}{c}
   \Phi_{2} (t, x) \\ \Phi_{3} (t,x) \\ \Phi_{1} (t,x)   
 \end{array}
 \right) ,  \ \Psi' := \lim_{x\rightarrow \xi_{\rm I}^-} \left(
 \begin{array}{c}
   \partial_x \Phi_{2} (t, x) \\ \partial_x \Phi_{3} (t,x) \\ \partial_x \Phi_{1} (t,x)   
 \end{array}
 \right) ,
\end{equation}
and for the Y-junction $(x_{4}x_{2}x_{3})$ (node $\rm I\!I$), we should take 
\begin{equation}
\label{eq:def-Psi2}
 \Psi := \lim_{x\rightarrow \xi_{\rm I\!I}^+} \left(
 \begin{array}{c}
   \Phi_{2} (t, x) \\ \Phi_{3} (t,x) \\ \Phi_{4} (t,x)   
 \end{array}
 \right) , \ \Psi' := \lim_{x\rightarrow \xi_{\rm I\!I}^+} \left(
 \begin{array}{c}
   \partial_x \Phi_{2} (t, x) \\ \partial_{x} \Phi_{3} (t,x) \\ \partial_x \Phi_{4} (t,x)   
 \end{array}
 \right) .
\end{equation}
Note that the above ordering of the axes in $\Psi$ and $\Psi'$
is different from that in the previous work \cite{fknt}.
The present ordering is useful to deal with 
external magnetic effects as seen below.
Equation (\ref{eq:jc3}) is equivalently expressed as \cite{cft} 
\begin{equation}
\label{eq:jc4}
 \left| \Psi - i L_{0} \Psi' \right|
  =  \left| \Psi + i L_{0} \Psi' \right| ,
\end{equation}
where $L_{0} (\in \mathbb{R})$ is a nonzero,  arbitrary constant 
with dimension of length, 
which can be regarded as a gauge freedom \cite{fknt} and, 
therefore, does not appear in physical quantities \cite{cft}. 
Equation (\ref{eq:jc4}) means that 
$\Psi - i L_{0} \Psi' $ is connected to $\Psi + i L_{0} \Psi' $ 
via a unitary transformation.
Thus, we obtain the junction condition \cite{cft} 
\begin{equation}
\label{eq:jc5}
 (U-I_{3} ) \Psi + i L_{0} (U+I_{3} ) \Psi' =  0,
\end{equation}
where $I_{3}$ is the $3\times 3$ identity matrix, 
and $U$ is a $3\times 3$ unitary matrix.
Therefore, the junction condition (\ref{eq:jc5}) is characterized by the 
unitary matrix  $U \in {\rm U}(3)$.

Let us also discuss a parametrization of the unitary matrix $U$. 
Based on the discussion in \cite{fknt}, 
we adopt the following parametrization. On the Y-junction
$(x_{1}x_{2}x_{3})$ (node I), we take 
\begin{equation}
\label{eq:u1}
 U=VDV^{\dagger} ,
\end{equation} 
where 
\begin{equation}
\label{eq:u2}
 D:={\rm diag} \left( 
  e^{i\theta_{(1)}} , e^{i\theta_{(2)}} , e^{i \theta_{(3)}} 
 \right),
\end{equation}
and 
\begin{equation}
\label{eq:u3}
 V:=e^{i\alpha \lambda_{3}} e^{i\beta \lambda_{2}}
     e^{i\gamma \lambda_{3}} e^{i\delta \lambda_{5}}
     e^{i a \lambda_{3}} e^{i b \lambda_{2}}.
\end{equation} 
Here,  $\theta_{(1)}$, $\theta_{(2)}$, $\theta_{(3)}$, 
$\alpha$, $\beta$, $\gamma$, $\delta$, $a$, $b$ $\in \mathbb{R}$,
and  $\lambda_{2}, \lambda_{3}$, and $\lambda_{5}$ are the Gell-Mann Matrices
(see \cite{fknt} in detail).
In the same way, on the Y-junction $(x_{4}x_{2}x_{3})$ (node I$\!$I),
we take
\begin{equation}
\label{eq:ut1}
 U=\tilde{V} \tilde{D} \tilde{V}^{\dagger} ,
\end{equation} 
where 
\begin{equation}
\label{eq:ut2}
 \tilde{D}:={\rm diag} \left( 
  e^{i\tilde{\theta}_{(1)}} , e^{i \tilde{\theta}_{(2)}} , e^{i \tilde{\theta}_{(3)}} 
 \right),
\end{equation}
and 
\begin{equation}
\label{eq:ut3}
 \tilde{V}:=e^{i\tilde{\alpha} \lambda_{3}} e^{i\tilde{\beta} \lambda_{2}}
     e^{i\tilde{\gamma} \lambda_{3}} e^{i\tilde{\delta} \lambda_{5}}
     e^{i \tilde{a} \lambda_{3}} e^{i \tilde{b} \lambda_{2}} .
\end{equation} 
Here $\tilde{\theta}_{(1)}$, $\tilde{\theta}_{(2)}$, $\tilde{\theta}_{(3)}$, 
$\tilde{\alpha}$, $\tilde{\beta}$, $\tilde{\gamma}$, $\tilde{\delta}$, 
$\tilde{a}$, $\tilde{b}$ $\in \mathbb{R}$.
Therefore, the junction condition  (\ref{eq:jc5}) with
the unitary matrix $U$ is characterized by the nine real parameters 
($\theta_{(1)}$, $\theta_{(2)}$, $\theta_{(3)}$, 
$\alpha$, $\beta$, $\gamma$, $\delta$, $a$, $b$),
or ($\tilde{\theta}_{(1)}$, $\tilde{\theta}_{(2)}$, $\tilde{\theta}_{(3)}$, 
$\tilde{\alpha}$, $\tilde{\beta}$, $\tilde{\gamma}$, $\tilde{\delta}$, 
$\tilde{a}$, $\tilde{b}$).

\subsection{Scattering matrices}

\subsubsection{The scattering matrix for a single Y-junction} 

We consider a quantum state on the coordinate system 
for the node I (see Fig.~\ref{fig2}(a)).
We assume that incoming waves and outgoing waves are provided by 
$\phi_{i} e^{ikx_{i}}$  and $\psi_{i} e^{-ikx_{i}}$ 
($\phi_{i}, \psi_{i} \in\mathbb{C} $ 
and $i=1, 2, 3 $ ), respectively. Then we have
\begin{equation}
\label{eq:wf}
 \Phi_{i} (t, x_{i} ) 
 = e^{-i\frac{\cal E}{\hbar} t} \left( 
   \phi_{i} e^{ikx_{i}} + \psi_{i} e^{-ikx_{i}} \right)  ,
\end{equation}
where ${\cal E} := \hbar^2 k^2 /2m$.
From this expression, we derive
\begin{equation}
\label{eq:Psi}
 \Psi 
 = e^{-i\frac{\cal E}{\hbar} t} \left\{ 
   e^{ik\xi_{\rm I}}\left( 
   \begin{array}{c}
     \phi_{2} \\ \phi_{3} \\ \phi_{1} 
   \end{array}
   \right)  + e^{-ik\xi_{\rm I}} \left( 
   \begin{array}{c}
     \psi_{2} \\ \psi_{3} \\ \psi_{1} 
   \end{array}
   \right) \right\}  ,
\end{equation}
\begin{equation}
\label{eq:Psid}
 \Psi' 
 = e^{-i\frac{\cal E}{\hbar} t} \left\{ 
   ik e^{ik\xi_{\rm I}}\left( 
   \begin{array}{c}
     \phi_{2} \\ \phi_{3} \\ \phi_{1} 
   \end{array}
   \right)  -ik e^{-ik\xi_{\rm I}} \left( 
   \begin{array}{c}
     \psi_{2} \\ \psi_{3} \\ \psi_{1} 
   \end{array}
   \right) \right\}  .
\end{equation}
By substituting Eqs.~(\ref{eq:Psi}) and (\ref{eq:Psid}) 
into Eq.~(\ref{eq:jc5}), 
we can define the $S$-matrix $S_{\rm I}$ by the equation
 \begin{equation}
\label{eq:s1}
 \left( 
   \begin{array}{c}
     \psi_{2} \\ \psi_{3} \\ \psi_{1} 
   \end{array}
   \right)
 = S_{\rm I} \left( 
   \begin{array}{c}
     \phi_{2} \\ \phi_{3} \\ \phi_{1} 
   \end{array}
   \right)    .
\end{equation}
Using Eqs.~(\ref{eq:u1}), (\ref{eq:u2}), and (\ref{eq:u3}), 
we derive
\begin{equation}
\label{eq:si}
 S_{\rm I} = e^{2ik\xi_{\rm I}} VS_{\rm (0)I} V^{\dagger} ,
\end{equation}
where
\begin{equation}
 S_{\rm (0)I} := {\rm diag}  \left( 
   \frac{ikL_{(1)}+1}{ikL_{(1)}-1}, \frac{ikL_{(2)}+1}{ikL_{(2)}-1}, 
 \frac{ikL_{(3)}+1}{ikL_{(3)}-1}
 \right) .
\end{equation}
Here we define
\begin{equation}
 L_{(i)} := L_{0} \cot \frac{\theta_{(i)}}{2} .
\end{equation}
When we write $S_{\rm I}$ in the form
\begin{equation}
 S_{\rm I} = \left( 
 \begin{array}{ccc}
   s_{22} & s_{23}  & s_{21} \\ 
   s_{32} & s_{33}  & s_{31} \\
   s_{12} & s_{13}  & s_{11} \\  
 \end{array} \right) ,
\end{equation}
then $s_{ii}$ represents the probability amplitude
for reflection from the $x_{i}$-axis to the $x_{i}$-axis, while the 
$s_{ij}$ ($i\neq j$) represents the probability amplitude
for transmission from the $x_{j}$-axis to the $x_{i}$-axis.
\footnote{If we adopt the parameters
 $\xi_{\rm I}=0, \ \alpha = 0 , \ \beta =\frac{3\pi}{2}, \ \gamma=\pi, 
 \ \delta=\frac{\pi}{4} , 
 a=0, \theta_{(1)}=0 , \ \theta_{(2)}=\theta_{(3)}=\pi$, then
 we can regain the $S$-matrix used in \cite{bia,buttiker} 
 (see also \cite{fknt}).}

Next, we consider a quantum state on
the coordinate system for the node I$\!$I  (see Fig.~\ref{fig2}(b)).
We assume that incoming waves and outgoing waves are provided by 
$\psi_{i} e^{-ikx_{i}}$  and $\phi_{i} e^{ikx_{i}}$ 
($\psi_{i}, \phi_{i} \in\mathbb{C} $ 
and $i=2, 3, 4 $ ), respectively. Then we have
\begin{equation}
\label{eq:phi}
 \Phi_{i} (t, x_{i} ) 
 = e^{-i\frac{\cal E}{\hbar} t} \left( 
   \psi_{i} e^{-ikx_{i}} + \phi_{i} e^{ikx_{i}} \right)  .
\end{equation}
From this expression,  we derive
\begin{equation}
\label{eq:Psi2}
 \Psi 
 = e^{-i\frac{\cal E}{\hbar} t} \left\{ 
   e^{-ik\xi_{\rm I\!I}}\left( 
   \begin{array}{c}
     \psi_{2} \\ \psi_{3} \\ \psi_{4} 
   \end{array}
   \right)  + e^{ik\xi_{\rm I\!I}} \left( 
   \begin{array}{c}
     \phi_{2} \\ \phi_{3} \\ \phi_{4} 
   \end{array}
   \right) \right\}  ,
\end{equation}
\begin{equation}
\label{eq:Psid2}
 \Psi' 
 = e^{-i\frac{\cal E}{\hbar} t} \left\{ 
   -ik e^{-ik\xi_{\rm I\!I}}\left( 
   \begin{array}{c}
     \psi_{2} \\ \psi_{3} \\ \psi_{4} 
   \end{array}
   \right)  +ik e^{ik\xi_{\rm I\!I}} \left( 
   \begin{array}{c}
     \phi_{2} \\ \phi_{3} \\ \phi_{4} 
   \end{array}
   \right) \right\}  .
\end{equation}
By substituting Eqs.~(\ref{eq:Psi2}) and (\ref{eq:Psid2}) 
into Eq.~(\ref{eq:jc5}), 
we can define the $S$-matrix $S_{\rm I\!I}$  by the equation
 \begin{equation}
\label{eq:s2}
 \left( 
   \begin{array}{c}
     \phi_{2} \\ \phi_{3} \\ \phi_{4} 
   \end{array}
   \right)
 = S_{\rm I\!I} \left( 
   \begin{array}{c}
     \psi_{2} \\ \psi_{3} \\ \psi_{4} 
   \end{array}
   \right)    .
\end{equation}
Using Eqs.~(\ref{eq:ut1}), (\ref{eq:ut2}), and (\ref{eq:ut3}), 
we derive
\begin{equation}
\label{eq:sii}
 S_{\rm I\!I} = e^{-2ik\xi_{\rm I\!I}} \tilde{V} 
S_{\rm (0)I\!I} \tilde{V}^{\dagger} ,
\end{equation}
where
\begin{equation}
 S_{\rm (0)I\!I} := {\rm diag} \left( 
   \frac{ik\tilde{L}_{(1)}-1}{ik\tilde{L}_{(1)}+1}, 
 \frac{ik\tilde{L}_{(2)}-1}{ik\tilde{L}_{(2)}+1}, 
 \frac{ik\tilde{L}_{(3)}-1}{ik\tilde{L}_{(3)}+1} 
 \right) .
\end{equation}
Here 
\begin{equation}
 \tilde{L}_{(i)} := \tilde{L}_{0} \cot \frac{\tilde{\theta}_{(i)}}{2} ,
\end{equation}
where $\tilde{L}_{0}$ denotes the parameter at the node ${\rm I\!I}$.
We can write $S_{\rm I\!I}$ in the form
\begin{equation}
 S_{\rm I\!I} = \left( 
 \begin{array}{ccc}
   \tilde{s}_{22} & \tilde{s}_{23}  & \tilde{s}_{24} \\ 
   \tilde{s}_{32} & \tilde{s}_{33}  & \tilde{s}_{34} \\
   \tilde{s}_{42} & \tilde{s}_{43}  & \tilde{s}_{44} \\  
 \end{array} \right) ,
\end{equation}
in common with the node I.

We also show important relations which 
$S_{\rm I}$ and $S_{\rm I\!I}$ satisfy.
Since $S_{\rm I}$ and $S_{\rm I\!I}$ are both unitary, we have
\begin{equation}
\label{eq:si-sii-r1}
 S_{\rm I} S_{\rm I}^{\dagger} =S_{\rm I}^{\dagger} S_{\rm I} =I_{3} , 
\end{equation}
\begin{equation}
\label{eq:si-sii-r2}
 S_{\rm I\!I} S_{\rm I\!I}^{\dagger} =
 S_{\rm I\!I}^{\dagger} S_{\rm I\!I} =I_{3} .
\end{equation}
These are also expressed as 
\begin{equation}
\label{eq:sij-r1}
 \sum_{k=1}^{3} s_{ik} s_{jk}^{\ast} 
 =\sum_{k=1}^{3} s_{ki}^{\ast} s_{kj} = \delta_{ij} ,
\end{equation}
\begin{equation}
\label{eq:sij-r2}
 \sum_{k=2}^{4} \tilde{s}_{ik} \tilde{s}_{jk}^{\ast}
   =\sum_{k=2}^{4} \tilde{s}_{ki}^{\ast} \tilde{s}_{kj} = \delta_{ij} .
\end{equation}
These relations are important to derive simplified 
forms for probability amplitudes in the next section.

\subsubsection{The scattering matrix for the ring } 

\begin{center}
\begin{figure}[h]
  \includegraphics[width=.8 \textwidth]{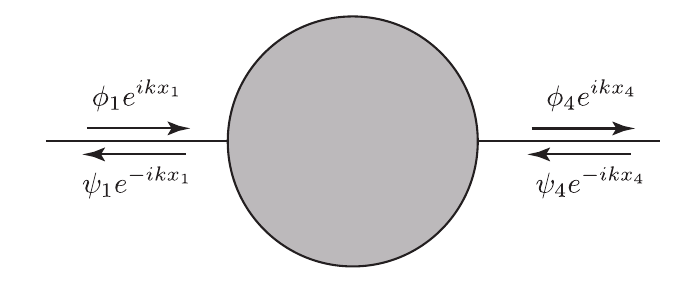}
  \caption{\label{fig3} Scattering by a ring system. 
   The gray circle includes the ring. }
\end{figure}
\end{center}

In this section, we consider the ring system shown in 
Fig.~\ref{fig1} as in Fig.~\ref{fig3}.
The $S$-matrix $S_{\rm R}$ for the ring system is defined by
\begin{equation}
\label{eq:ring-s-martix}
 \left( 
 \begin{array}{c}
  \psi_{1} \\ \phi_{4}  
 \end{array}\right) 
 = S_{\rm R}
  \left( 
 \begin{array}{c}
  \phi_{1} \\ \psi_{4}  
 \end{array}\right) ,
\end{equation} 
where $S_{\rm R}$ is a $2\times 2$ matrix. 
We derive $S_{\rm R}$ in terms of the components of $S_{\rm I}$
and $S_{\rm I\!I}$, i.e., $s_{ij}$ and $\tilde{s}_{ij}$. 
For this purpose, we define \(2\times 2\) matrices \(s\) and \(\tilde{s}\) as
\begin{equation}
 s:= \left( 
 \begin{array}{cc}
   s_{22} & s_{23} \\ s_{32} & s_{33} 
 \end{array}
 \right) , \quad 
 \tilde{s} := \left( 
 \begin{array}{cc}
   \tilde{s}_{22} & \tilde{s}_{23} \\ \tilde{s}_{32} & \tilde{s}_{33} 
 \end{array}
 \right) .
\end{equation}
By decomposing Eqs.~(\ref{eq:s1}) and (\ref{eq:s2})
into the components on $x_1$ and $x_4$ axes 
and those on $x_2$ and $x_3$ axes, we can obtain
the components of $S_{\rm R}$ as
\begin{eqnarray}
\label{eq:sr11}
 \left( S_{\rm R} \right)_{11} 
 &=&  s_{11} + \left( s_{12} \ \ s_{13} \right) 
  \left( I_{2} - \tilde{s}s \right)^{-1}
  \tilde{s} \left(
  \begin{array}{c}
    s_{21} \\ s_{31} 
  \end{array}
  \right)  , \\
\label{eq:sr12}
  \left( S_{\rm R} \right)_{12} 
  &= & \left( s_{12} \ \ s_{13} \right) 
  \left( I_{2} - \tilde{s}s \right)^{-1}
  \left(
  \begin{array}{c}
    \tilde{s}_{24} \\ \tilde{s}_{34} 
  \end{array}
  \right)  , \\
\label{eq:sr21}
  \left( S_{\rm R} \right)_{21} 
  &=&
  \left( \tilde{s}_{42} \ \ \tilde{s}_{43} \right) 
  \left( I_{2} - s\tilde{s} \right)^{-1}
  \left(
  \begin{array}{c}
    s_{21} \\ s_{31} 
  \end{array}
  \right)  , \\
\label{eq:sr22}
  \left( S_{\rm R} \right)_{22} 
  &= &  \tilde{s}_{44} + \left( \tilde{s}_{42} \ \ \tilde{s}_{43} \right) 
  \left( I_{2} - s\tilde{s} \right)^{-1}
  s \left(
  \begin{array}{c}
    \tilde{s}_{24} \\ \tilde{s}_{34} 
  \end{array}
  \right)  ,
\end{eqnarray}
where  $I_{2}$ is the $2\times 2$ identity matrix.
Here, we assume $|s_{ij}| <1$ and $|\tilde{s}_{ij}|<1$.
In this case,  $(I_{2} - s\tilde{s} )^{-1}$ and $(I_{2} - \tilde{s}s )^{-1}$ are given by
\begin{eqnarray}
 (I_{2} - s\tilde{s} )^{-1} 
 &=& \sum_{k=0}^{\infty} \left( s\tilde{s} \right)^{k} , \\ \nonumber \\
 (I_{2} - \tilde{s}s )^{-1} 
 &=& \sum_{k=0}^{\infty} \left( \tilde{s}s \right)^{k} . \\ \nonumber 
\end{eqnarray}
While the diagonal components of  $S_{\rm R}$ 
represent the probability amplitude for reflection, 
the non-diagonal components of $S_{\rm R}$
represent that for transmission.

\section{Localized states on the ring}
\label{sec: l_states}

We discuss localized states which could arise on the ring.
In these states, while $\Phi_{1}$ and $\Phi_{4}$ vanish
(i.e.,  $\Phi_{1}=\Phi_{4}=0$), the non-zero components
$\Phi_{2}$ and $\Phi_{3}$ are given by Eq.~(\ref{eq:wf}).
The junction conditions at node I 
and node I$\!$I are then written as
\begin{equation}
\label{eq:l_jc}
 M \left(
 \begin{array}{c}
    \phi_{2} \\ \psi_{2} \\ \phi_{3} \\ \psi_{3}
 \end{array} \right) =0 ,
\end{equation}
where
\begin{equation}
 M:= \left( 
 \begin{array}{cccc}
   V_{21}^{\ast} \kappa_{1}  e^{2ik\xi_{\rm I}} 
   &  V_{21}^{\ast} \kappa_{1}^{\ast}   &
      V_{31}^{\ast} \kappa_{1} e^{2ik\xi_{\rm I}} 
   &  V_{31}^{\ast} \kappa_{1}^{\ast}  \\
   V_{22}^{\ast} \kappa_{2}  e^{2ik\xi_{\rm I}} 
   &  V_{22}^{\ast} \kappa_{2}^\ast  &
      V_{32}^{\ast} \kappa_{2}  e^{2ik\xi_{\rm I}} 
   &  V_{32}^{\ast} \kappa_{2}^\ast  \\
   V_{23}^{\ast} \kappa_{3}  e^{2ik\xi_{\rm I}} 
   &  V_{23}^{\ast} \kappa_{3}^\ast  &
      V_{33}^{\ast} \kappa_{3}  e^{2ik\xi_{\rm I}} 
   &  V_{33}^{\ast} \kappa_{3}^\ast  \\
   \tilde{V}_{21}^{\ast} \tilde{\kappa}_{1}  e^{2ik\xi_{\rm I\!I}} 
 &  \tilde{V}_{21}^{\ast} \tilde{\kappa}_{1}^\ast   &
      \tilde{V}_{31}^{\ast} \tilde{\kappa}_{1}   e^{2ik\xi_{\rm I\!I}} 
 &  \tilde{V}_{31}^{\ast} \tilde{\kappa}_{1}^\ast   \\
   \tilde{V}_{22}^{\ast} \tilde{\kappa}_{2}   e^{2ik\xi_{\rm I\!I}} 
 &  \tilde{V}_{22}^{\ast} \tilde{\kappa}_{2}^\ast   &
      \tilde{V}_{32}^{\ast} \tilde{\kappa}_{2}   e^{2ik\xi_{\rm I\!I}} 
 &  \tilde{V}_{32}^{\ast} \tilde{\kappa}_{2}^\ast  \\
   \tilde{V}_{23}^{\ast} \tilde{\kappa}_{3}   e^{2ik\xi_{\rm I\!I}} 
 &  \tilde{V}_{23}^{\ast} \tilde{\kappa}_{3}^\ast  &
      \tilde{V}_{33}^{\ast} \tilde{\kappa}_{3}  e^{2ik\xi_{\rm I\!I}} 
 &  \tilde{V}_{33}^{\ast} \tilde{\kappa}_{3}^\ast  \\
 \end{array}
 \right) .
\end{equation}
Here $\kappa_i := 1+ik L_{(i)}$, $\tilde{\kappa}_i := 1+ik \tilde{L}_{(i)}$, and 
$V_{ij}$  and $\tilde{V}_{ij}$ denote the $(ij)$-components 
of the matrices $V$ and $\tilde{V}$, respectively. 
For the localized states, the normalization condition 
\begin{equation}
\label{eq:nc}
 \int_{\xi_{\rm I\!I}}^{\xi_{\rm I}} |\Phi_2 (t, x_{2})  |^2 dx_2
 +  \int_{\xi_{\rm I\!I}}^{\xi_{\rm I}} |\Phi_3 (t, x_{3}) |^2  dx_3 =1 
\end{equation}
should also be supplemented. 
The junction conditions given by Eq.~(\ref{eq:l_jc}) and 
the normalization condition (\ref{eq:nc}) provide seven equations 
for four unknown amplitudes $\phi_{2}, \psi_{2}, \phi_{3}$, and $\psi_{3}$.
Thus, this system of equations is over-determined.
Therefore, localized states on the ring are suppressed in general.

However, if the condition 
\begin{equation}
\label{eq:rank_M}
 {\rm rank} \ M \leq3 
\end{equation}
holds, then the number of equations for the junction conditions is 
reduced appropriately, and 
localized states appear on the ring.
If ${\rm rank} \ M <3$, 
the system of equations becomes under-determined.
Then some degrees of freedom remain and, therefore, 
degenerate states may appear.
It should also be noted that the symmetric ring 
in which the node on the Y-junction $(x_{1}x_{2}x_{3})$
has the same parameters as the node on 
the Y-junction $(x_{4}x_{2}x_{3})$
is special.  In the symmetric ring system, since
\begin{equation}
\label{eq:cnd_sr}
 V_{ij} = \tilde{V}_{ij}, \quad L_{(i)} =\tilde{L}_{(i)} ,
\end{equation}
the condition 
\begin{equation}
 {\rm rank} \ M=3 
\end{equation}
holds for any parameters about $U$ when
the wavenumber $k$ satisfies
\begin{equation}
\label{eq:cdt}
 e^{2ik (\xi_{\rm I} - \xi_{\rm I\!I})}=1 .
\end{equation}
Therefore, in the case of the symmetric ring,
there exist localized states
for any parameters about $U$.
The existence of such localized states may be suggested 
from the symmetry of the system (e.g., \cite{xgsy}).

Let us derive the wave function for the localized states in the symmetric ring systems.
When the condition Eq.~(\ref{eq:cdt}) holds, the wave function 
is determined by three independent components in Eq.~(\ref{eq:l_jc})
and the normalization condition (\ref{eq:nc}). 
From these equations, we can obtain the solution
\begin{eqnarray}
\label{eq:wf-lc1}
 \varphi_2 (x_{2}) 
 & = &\frac{1}{\cal N} \left\{  {\cal C}_{2} \sin k (x_{2}- \xi_{\rm I\!I} ) 
     +  {\cal D}_{2} \cos k (x_{2}- \xi_{\rm I\!I} )  \right\} ,
 \\
\label{eq:wf-lc2}
  \varphi_3 (x_{3}) 
 & = & \frac{1}{\cal N} \left\{   {\cal C}_{3} \sin k (x_{3}- \xi_{\rm I\!I} ) 
         +  {\cal D}_{3} \cos k (x_{3}- \xi_{\rm I\!I} ) \right\} ,
\end{eqnarray}
where
\begin{eqnarray}
 {\cal C}_{2}
   & = &  V_{21}^{\ast} V_{31} kL_{(1)} 
           + V_{22}^{\ast} V_{32} kL_{(2)} + V_{23}^{\ast} V_{33} kL_{(3)}  , \\
 {\cal D}_{2}
   & = &  V_{21}^{\ast} V_{31} k^2L_{(2)} L_{(3)} 
         + V_{22}^{\ast} V_{32} k^2 L_{(3)} L_{(1)} + V_{23}^{\ast} V_{33} k^2 L_{(1)} L_{(2)}  , \\
 {\cal C}_{3}
   & = &  V_{11}^{\ast} V_{31} kL_{(1)} 
           + V_{12}^{\ast} V_{32} kL_{(2)} + V_{13}^{\ast} V_{33} kL_{(3)}  , \\
 {\cal D}_{3}
   & = &  V_{11}^{\ast} V_{31} k^2L_{(2)} L_{(3)} 
         + V_{12}^{\ast} V_{32} k^2 L_{(3)} L_{(1)} + V_{13}^{\ast} V_{33} k^2 L_{(1)} L_{(2)}  .
\end{eqnarray}
Here we define
\begin{equation}
 {\cal N}:= \sqrt{ \frac{1}{2} (\xi_{\rm I}- \xi_{\rm I\!I} )
 \left( \left| {\cal C}_{2} \right|^2 +  \left| {\cal D}_{2} \right|^2
      + \left| {\cal C}_{3} \right|^2 +\left| {\cal D}_{3} \right|^2 \right) } .
\end{equation}
From the above result, we find that the wave function can generally  take non-zero values
at the nodes, i.e., $\varphi_{2} (\xi_{\rm I}) = \pm {\cal D}_2/{\cal N}$, 
$\varphi_{2} (\xi_{\rm I\!I}) = {\cal D}_2 /{\cal N}$, $\varphi_{3} (\xi_{\rm I}) = \pm {\cal D}_3 /{\cal N}$, 
and $\varphi_{3} (\xi_{\rm I\!I}) = {\cal D}_3 /{\cal N}$.
Consequently, when the condition (\ref{eq:cdt}) holds, 
the localized states given by Eqs.~(\ref{eq:wf-lc1}) and (\ref{eq:wf-lc2})
appear on the symmetric ring.

\section{Transmission probability in a symmetric ring system}
\label{sec:trans}

\begin{center}
\begin{figure}[h]
  \includegraphics[width=.8 \textwidth]{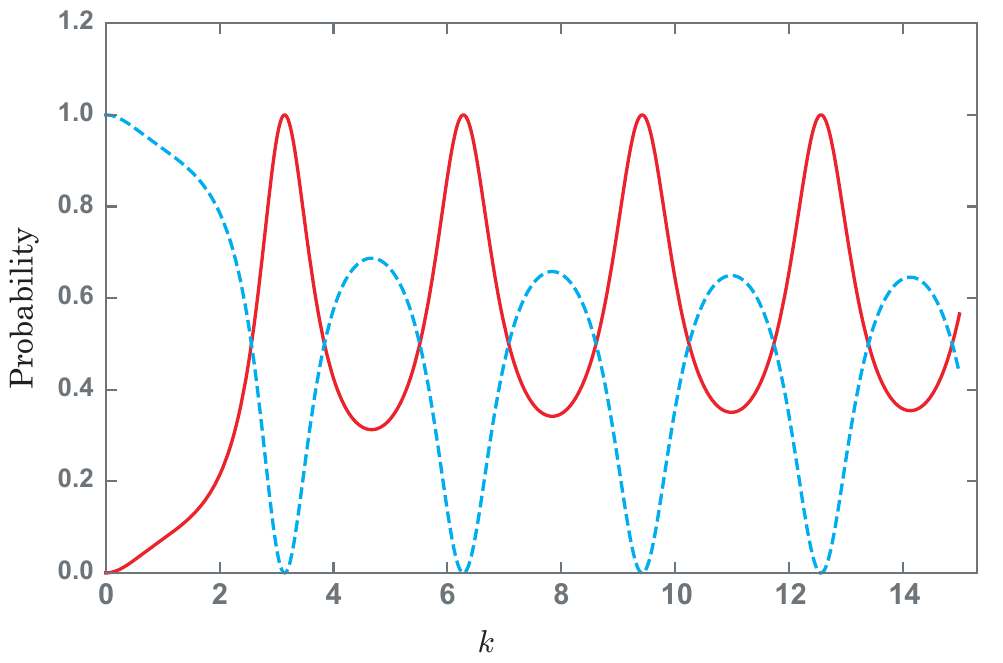}
  \caption{\label{fig4} Probabilities vs $k$ in the case of 
  $\alpha =\gamma=a =0$, $\beta = \delta=b=\frac{\pi}{4}$, $L_{(1)}=0$, 
  and $L_{(2)}=L_{(3)}=1$. Here we adopt $\xi_{\rm I}=1$ and $\xi_{\rm I\!I}=0$. 
  The red solid curve denotes the transmission probability, and the blue 
  dashed curve denotes the reflection probability.}
\end{figure}
\end{center}

In the last section, we have seen that 
the symmetric ring has remarkable feature that 
the localized states inevitably exist.
Hence, we focus on the symmetric ring systems 
also in the framework of scattering problems.
Let us assume 
\begin{equation}
\label{eq:pw}
 \phi_{1} =1, \  \psi_{1}=R, \ \phi_{4}=T , \ \psi_{4}=0. 
\end{equation}
where $R$ is the amplitude for reflection, and $T$ is the 
amplitude for transmission.
%
From Eq.~(\ref{eq:ring-s-martix}) and Eqs.~(\ref{eq:sr11})--(\ref{eq:sr22}), 
we can obtain 
\begin{eqnarray}
\label{eq:R}
 R & = & s_{11} + \left( s_{12} \ \  s_{13} \right)
 \tilde{s} \left( I_{2} -s\tilde{s} \right)^{-1} 
 \left( 
  \begin{array}{c}
    s_{21} \\ s_{31}
  \end{array}
 \right)  , \\
\label{eq:T}
 T & = & \left( \tilde{s}_{42} \  \ \tilde{s}_{43} \right)
 \left( I_{2} -s\tilde{s} \right)^{-1} 
 \left( 
  \begin{array}{c}
    s_{21} \\ s_{31}
  \end{array}
 \right)  .
\end{eqnarray}
When Eq.~(\ref{eq:cnd_sr}) holds,  we have
\begin{equation}
\label{eq:si-sii-r3}
 S_{\rm I} S_{\rm I\!I} =S_{\rm I\!I} S_{\rm I} 
 = e^{2ik \left( \xi_{\rm I} - \xi_{\rm I\!I} \right)} I_{3} .
\end{equation}
Thus we derive
\begin{equation}
\label{eq:si-sii-r4}
 S_{\rm I\!I} = e^{2ik \left( \xi_{\rm I} - \xi_{\rm I\!I} \right)} 
 S_{\rm I}^{\dagger} ,
\end{equation}
i.e.,
\begin{eqnarray}
\label{eq:si-sii-r4c}
 \tilde{s}_{ij} = e^{2ik \left( \xi_{\rm I} - \xi_{\rm I\!I} \right)} 
 s_{ji}^{\ast} , \\
 \tilde{s}_{4j} = e^{2ik \left( \xi_{\rm I} - \xi_{\rm I\!I} \right)} 
 s_{j1}^{\ast} ,  \\ 
 \tilde{s}_{i4} = e^{2ik \left( \xi_{\rm I} - \xi_{\rm I\!I} \right)} 
 s_{1i}^{\ast} , \\ 
\label{eq:si-sii-r4c2}
 \tilde{s}_{44} = e^{2ik \left( \xi_{\rm I} - \xi_{\rm I\!I} \right)} 
 s_{11}^{\ast} , 
\end{eqnarray}
where $i, j=2$ or $3$.
By using these equations and Eq.~(\ref{eq:sij-r1}), we can derive
\begin{equation}
 s\tilde{s} = e^{2ik \left( \xi_{\rm I} - \xi_{\rm I\!I} \right)} 
 \left( 
 \begin{array}{cc}
 |s_{22} |^2 + | s_{23} |^2 & -s_{21}s_{31}^{\ast}   \\
 -s_{21}^{\ast} s_{31} &  |s_{32} |^2 + | s_{33} |^2
 \end{array}
 \right) .
\end{equation}
Then we have
\begin{equation}
\label{eq:sts}
 s\tilde{s} 
 \left( 
 \begin{array}{c}
  s_{21}  \\ s_{31}
 \end{array}
 \right) =  e^{2ik \left( \xi_{\rm I} - \xi_{\rm I\!I} \right)}
  \left| s_{11} \right|^2  \left( 
 \begin{array}{c}
  s_{21}  \\ s_{31}
 \end{array}
 \right) ,
\end{equation}
and 
\begin{eqnarray}
 \left( I_{2} - s\tilde{s} \right)^{-1} 
 \left( 
 \begin{array}{c}
  s_{21}  \\ s_{31}
 \end{array}
 \right) &= & \left\{ I_{2} +  s\tilde{s}  + \left(  s\tilde{s}  \right) 
       + \cdots \right\} \left( 
 \begin{array}{c}
  s_{21}  \\ s_{31}
 \end{array}
 \right) \nonumber \\
 &= & \frac{1}{1- e^{2ik \left( \xi_{\rm I} - \xi_{\rm I\!I} \right)}
  \left| s_{11} \right|^2}  \left( 
 \begin{array}{c}
  s_{21}  \\ s_{31}
 \end{array}
 \right) .
\end{eqnarray}
Therefore, we obtain 
\begin{eqnarray}
\label{eq:sR}
 R & = & \frac{s_{11} 
 \left( 1- e^{2ik \left( \xi_{\rm I} - \xi_{\rm I\!I} 
 \right)} \right)}
 {1- e^{2ik \left( \xi_{\rm I} - \xi_{\rm I\!I} \right)}
  \left| s_{11} \right|^2} , \\
\label{eq:sT}
 T & = &  
 \frac{e^{2ik \left( \xi_{\rm I} - \xi_{\rm I\!I} \right)} 
 \left( 1-|s_{11}|^2 \right) }
 {1- e^{2ik \left( \xi_{\rm I} - \xi_{\rm I\!I} \right)}
  \left| s_{11} \right|^2}   .
\end{eqnarray}
In Fig.~\ref{fig4}, we show transmission and reflection probabilities as
a function of the wavenumber $k$ as an example. From this figure, 
we can find that the perfect transmission occurs 
when the condition in Eq.~(\ref{eq:cdt}) holds. 
Thus, it should be emphasized that
when the same condition as Eq.~(\ref{eq:cdt}) holds, 
the perfect transmission, which is given by $R=0$, occurs simultaneously 
with the appearance of the localized states in the symmetric ring.

\section{Disturbance to the symmetric ring conditions}
\label{sec:md}

\subsection{Formulation for magnetic effects}

In this section, we investigate effects of disturbance 
to the symmetric ring.
As a most typical example, we consider effects of 
external magnetic fields on the quantum states.
When we consider a charged particle on the ring and 
an external magnetic flux penetrating 
the ring, we should replace the partial derivative $\partial_{x}$ 
with the covariant derivative $D_{x}:=\partial_{x} -i\frac{e}{\hbar c} A_{x}$, 
where $e$ is an electric charge, and 
$A_{x}$ denotes the vector potential of magnetic fields,
in Eqs.~(\ref{eq:se}), (\ref{eq:def-Psi1}) and (\ref{eq:def-Psi2}).
Let us also assume non-vanishing magnetic flux confined inside the ring. 
Since magnetic field on the wires vanishes, the magnetic vector potential 
is provided by pure gauge. The wave function in the presence of 
the magnetic flux is given by \cite{LL}
\begin{eqnarray}
 \Phi_{i}^{\rm (M)} (x_{i} ) & = & 
     e^{i\frac{e}{\hbar c}\chi_i  (x_{i})} \Phi_{i} (x_{i} )    \quad (i=1, 2, 3, 4) ,
\end{eqnarray}
where $\Phi_i$ is the wave function in the absence of 
the magnetic flux, which is given by Eq.~(\ref{eq:wf}), and 
\begin{eqnarray}
 \chi_i  (x) & : = & \int^{x}_{\eta_{i}} A_{i} dx_{i}   \quad (i=1, 2, 3, 4) .
\end{eqnarray}
Here $A_{i}$ denotes the tangential component of three-dimensional vector potential $\bi{A}$
along the $x_{i}$ axis, and $\eta_{i}$ is an arbitrary constant. 
We now take 
\begin{equation}
 \eta_{1} = \eta_{2} = \eta_{3} = \xi_{\rm I}, \quad  
 \eta_{4} = \xi_{\rm I\!I} .
\end{equation}
Then we have
\begin{eqnarray}
 \chi_1  (\xi_{\rm I})=\chi_2  (\xi_{\rm I})=\chi_3  (\xi_{\rm I})=0 , \quad 
 \chi_4  (\xi_{\rm I\!I})=0 .
\end{eqnarray}
Under this assumption,  we derive
\begin{eqnarray}
 \Psi^{\rm (M)} (\xi_{\rm I}) = \Psi (\xi_{\rm I}) , \quad 
 \Psi^{\prime \rm (M)} (\xi_{\rm I}) = \Psi^{\prime}(\xi_{\rm I})  , \\
 \Psi^{\rm (M)} (\xi_{\rm I\!I}) = P^{\dagger} \Psi (\xi_{\rm I\!I}) , \quad 
 \Psi^{\prime \rm (M)} (\xi_{\rm I\!I}) = P^{\dagger} \Psi^{\prime} (\xi_{\rm I\!I}) ,
\end{eqnarray}
where $\Psi$ and $\Psi'$ are given by Eqs.~(\ref{eq:Psi}) and (\ref{eq:Psid}), and 
we define
\begin{equation}
 P:={\rm diag} \left( e^{-i\frac{e}{\hbar c}\chi_2 (\xi_{\rm I\!I})},
 e^{-i\frac{e}{\hbar c}\chi_3 (\xi_{\rm I\!I})} , 1 \right)  .
\end{equation}
Hence  the junction conditions in the presence of the magnetic flux 
are replaced with
\begin{equation}
\label{eq:jc6}
 (U -I_{3} ) \Psi (\xi_{\rm I}) + i L_{0} (U +I_{3} ) \Psi' (\xi_{\rm I}) =  0 ,
\end{equation}
\begin{equation}
\label{eq:jc7}
 (PUP^{\dagger} -I_{3} ) \Psi (\xi_{\rm I\!I}) 
 + i L_{0} (PUP^{\dagger} +I_{3} ) \Psi' (\xi_{\rm I\!I}) =  0 .
\end{equation}
Since $PUP^{\dagger} = P\tilde{V} \tilde{D} ( P\tilde{V} )^{\dagger}$
at the node I$\!$I, the effect of  the magnetic flux is expressed by 
\begin{equation}
  \tilde{V} \longrightarrow \tilde{V}^{(\rm M)} = P\tilde{V} .
\end{equation}
Note that the magnetic flux $\phi_{0}$ penetrating the ring is given by
\begin{equation}
 \phi_{0} = \chi_3 (\xi_{\rm I\!I}) - \chi_2 (\xi_{\rm I\!I})
 = \oint_{\rm Ring}  \bi{A} \cdot d\bi{r} .
\end{equation}
When we consider a symmetric configuration for the ring arms, we have
\begin{equation}
 \chi_3 (\xi_{\rm I\!I}) =  - \chi_2 (\xi_{\rm I\!I}).
\end{equation}
Then we derive
\begin{equation}
 \chi_2 (\xi_{\rm I\!I}) = -\frac{1}{2} \phi_{0}, \quad 
 \chi_3 (\xi_{\rm I\!I}) = \frac{1}{2} \phi_{0}.
\end{equation}
Hence we obtain
\begin{equation}
 P:={\rm diag} \left( e^{i\frac{e\phi_{0}}{2 \hbar c}},
 e^{-i\frac{e\phi_{0}}{2 \hbar c}} , 1 \right)  .
\end{equation}
This expression leads to 
\begin{eqnarray}
  \tilde{V} \longrightarrow \tilde{V}^{(\rm M)} 
  & = & P\tilde{V} \nonumber  \\
  & = & e^{i\tilde{\alpha}^{\rm (M)} \lambda_{3}} e^{i\tilde{\beta} \lambda_{2}}
     e^{i\tilde{\gamma} \lambda_{3}} e^{i\tilde{\delta} \lambda_{5}}
     e^{i \tilde{a} \lambda_{3}} e^{i \tilde{b} \lambda_{2}} ,
\end{eqnarray}
where
\begin{equation}
 \tilde{\alpha}^{\rm (M)} := \tilde{\alpha} + \frac{e\phi_{0}}{2 \hbar c} .
\end{equation}
Therefore, when we assume the symmetric configuration for the ring arms,
the effect of the magnetic flux can be 
expressed by the modulation of the parameter $\tilde{\alpha}$.

\subsection{Effects of the magnetic disturbance on the transmission probability}

We investigate the effects of the magnetic disturbance on the transmission probability.
For the symmetric ring systems which the magnetic flux penetrates,
the scattering matrix at the node I$\!$I becomes
\begin{equation}
 S_{\rm I\!I} = e^{2ik \left(\xi_{\rm I} - \xi_{\rm I\!I}\right)} 
e^{i\frac{e\phi_0}{2 \hbar c} \lambda_{3}}  
S_{\rm I}^{\dagger} e^{-i\frac{e\phi_0}{2 \hbar c} \lambda_{3}}.
\end{equation}
From Eqs.~(\ref{eq:R}) and (\ref{eq:T}), then we derive
\begin{eqnarray}
\label{eq:me-R}
 R & = & \frac{1}{\Delta}  \left[ s_{11} 
 \left( 1- e^{2ik \left( \xi_{\rm I} - \xi_{\rm I\!I}  \right)} \right)^2
 + 2i e^{2ik \left( \xi_{\rm I} - \xi_{\rm I\!I}  \right)}  \sin \frac{e \phi_{0}}{2\hbar c}
 \right. \nonumber \\
& &  \times \left.
\left\{   e^{-i\frac{e \phi_{0}}{2 \hbar c}} 
  s_{23}^{\ast} \left( s_{11} s_{23} - s_{13}s_{21} \right) 
  - e^{i\frac{e \phi_{0}}{2 \hbar c}} 
  s_{32}^{\ast} \left( s_{11} s_{32} - s_{12}s_{31} \right)  \right\}  \right] \\
\label{eq:me-T}
 T & = &  
 \frac{e^{2ik \left( \xi_{\rm I} - \xi_{\rm I\!I}  \right)}}{\Delta}  \left[ 
\left( 1 -  |s_{11}|^2 \right)
 \left( 1- e^{2ik \left( \xi_{\rm I} - \xi_{\rm I\!I}  \right)} \right)
 \left( 1+ 2i  e^{i\frac{e \phi_{0}}{4 \hbar c}} \sin \frac{e \phi_{0}}{4\hbar c} \right)
 \right. \nonumber \\
&& \left.  -  2i   \sin \frac{e \phi_{0}}{2\hbar c} 
 \left( |s_{21}|^2 - e^{2ik \left( \xi_{\rm I} - \xi_{\rm I\!I}  \right)} |s_{12}|^2 \right)  
  \right] ,
\end{eqnarray}
where 
\begin{eqnarray}
 \Delta & =&  \left( 1- e^{2ik \left( \xi_{\rm I} - \xi_{\rm I\!I}  \right)} |s_{11}|^2 \right)
        \left( 1- e^{2ik \left( \xi_{\rm I} - \xi_{\rm I\!I}  \right)} \right) \nonumber \\ 
&& +2i e^{2ik \left( \xi_{\rm I} - \xi_{\rm I\!I}  \right)} \sin \frac{e \phi_{0}}{2\hbar c} 
     \left(e^{-i\frac{e \phi_{0}}{2 \hbar c}} |s_{23}|^2  - e^{i\frac{e \phi_{0}}{2 \hbar c}} |s_{32}|^2 \right).
\end{eqnarray}
In the limit of vanishing magnetic flux, i.e., $\phi_{0}\rightarrow 0$, we retrieve
the results in Eqs.~(\ref{eq:sR}) and (\ref{eq:sT}).
Equations (\ref{eq:me-R}) and (\ref{eq:me-T}) provide 
the general expressions of the amplitudes for the reflection and transmission
in the symmetric ring system which magnetic flux penetrates.
Note that $s_{ij}$ in the above equations are functions of the nine parameters
of the Y-junction.

\begin{center}
\begin{figure}[h]
  \includegraphics[width=.5 \textwidth]{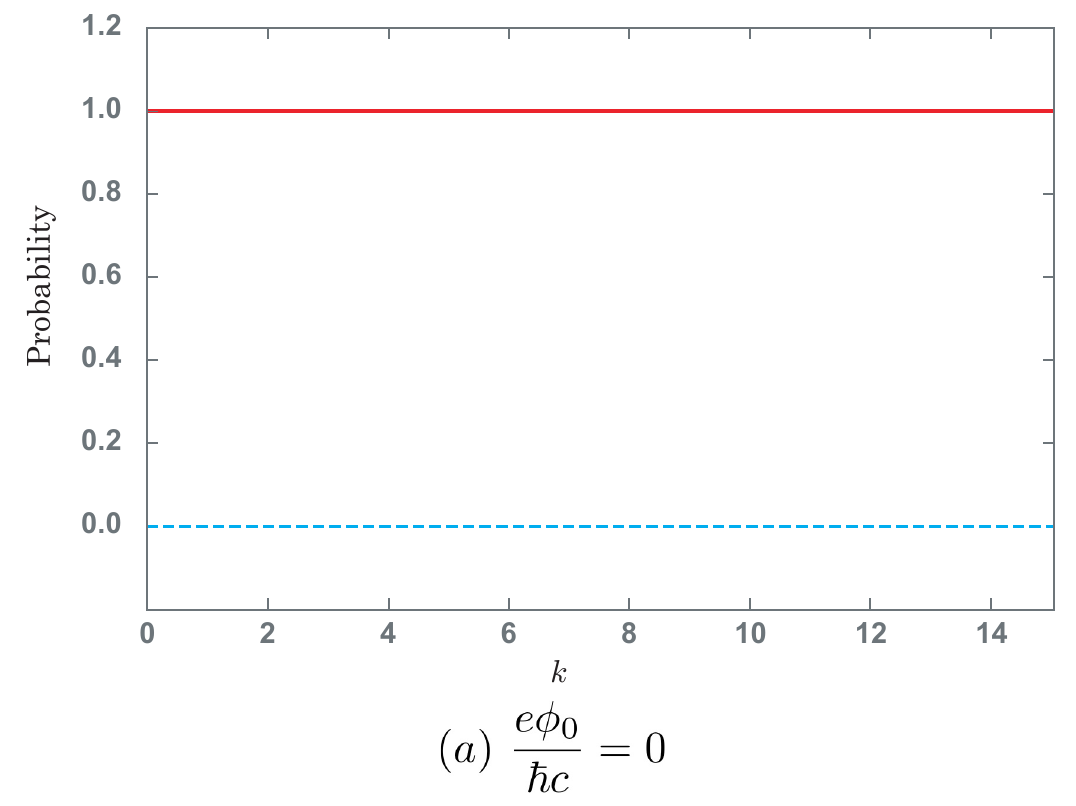}
  \includegraphics[width=.5 \textwidth]{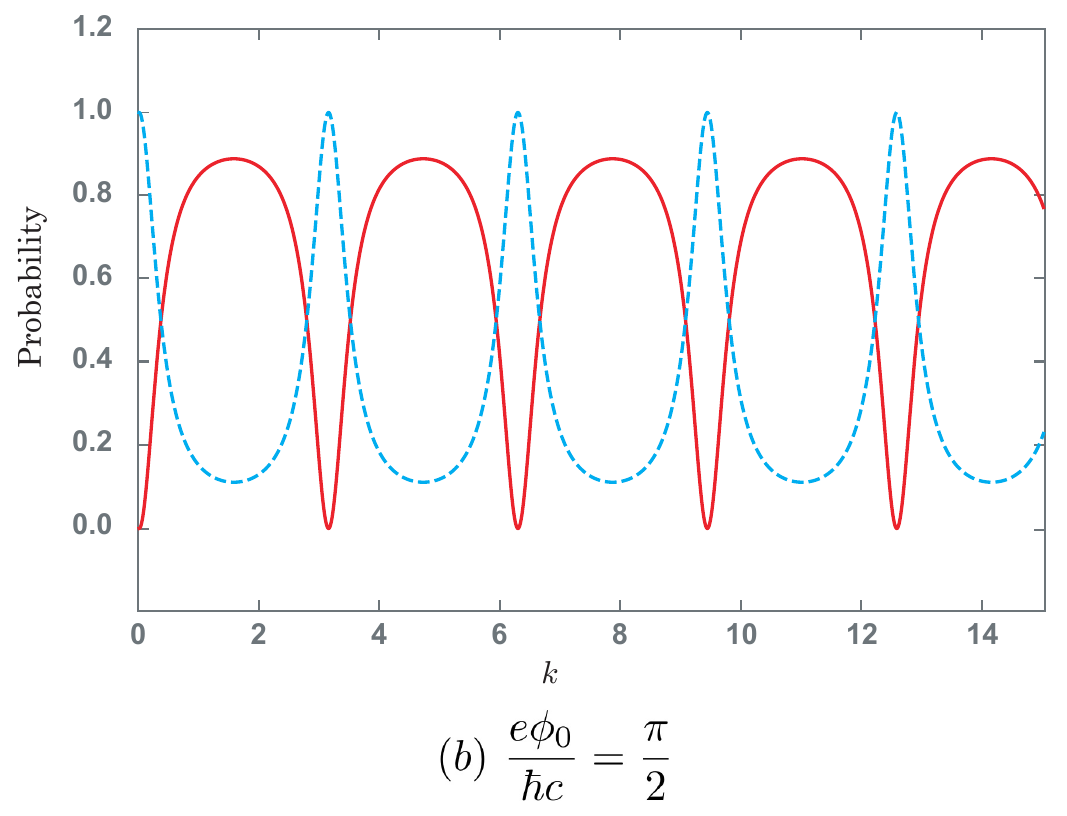}
  \includegraphics[width=.5 \textwidth]{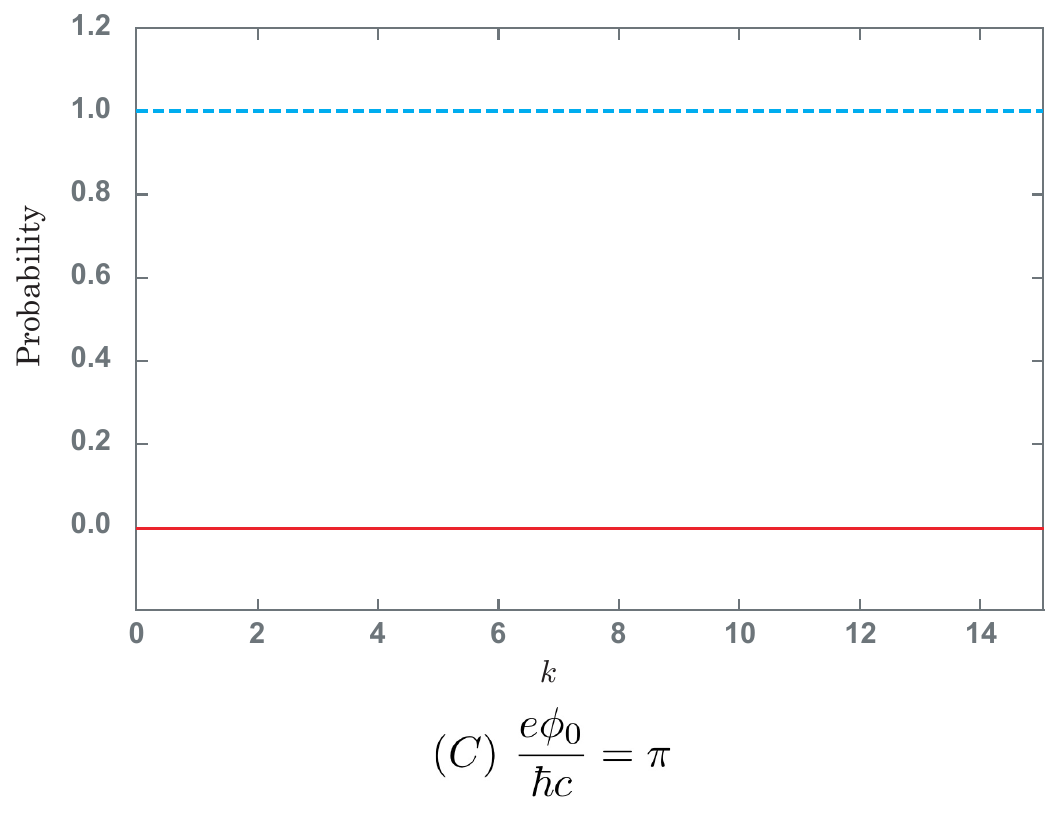}
  \caption{\label{fig5} Probabilities vs $k$ 
   in the cases of (a) $\frac{e\phi_{0}}{\hbar c}=0$, (b) $\frac{e\phi_{0}}{\hbar c}=\frac{\pi}{2}$,
   and (c) $\frac{e\phi_{0}}{\hbar c}=\pi$.
   Here we adopt $\alpha =\gamma=a =0$, $\beta = \delta=b=\frac{\pi}{4}$, $L_{(1)}=L_{(2)}=0$, 
   $L_{(3)}=\infty$, $\xi_{\rm I}=1$ and $\xi_{\rm I\!I}=0$. 
   The red solid curve denotes the transmission probability, and the blue 
   dashed curve denotes the reflection probability.
   Note that when $\frac{e\phi_{0}}{\hbar c}=2n\pi$ ($n$ is an integer), 
   we regain the same result as in the case of  $\frac{e\phi_{0}}{\hbar c}=0$. }
\end{figure}
\end{center}

\begin{center}
\begin{figure}[h]
  \includegraphics[width=.8 \textwidth]{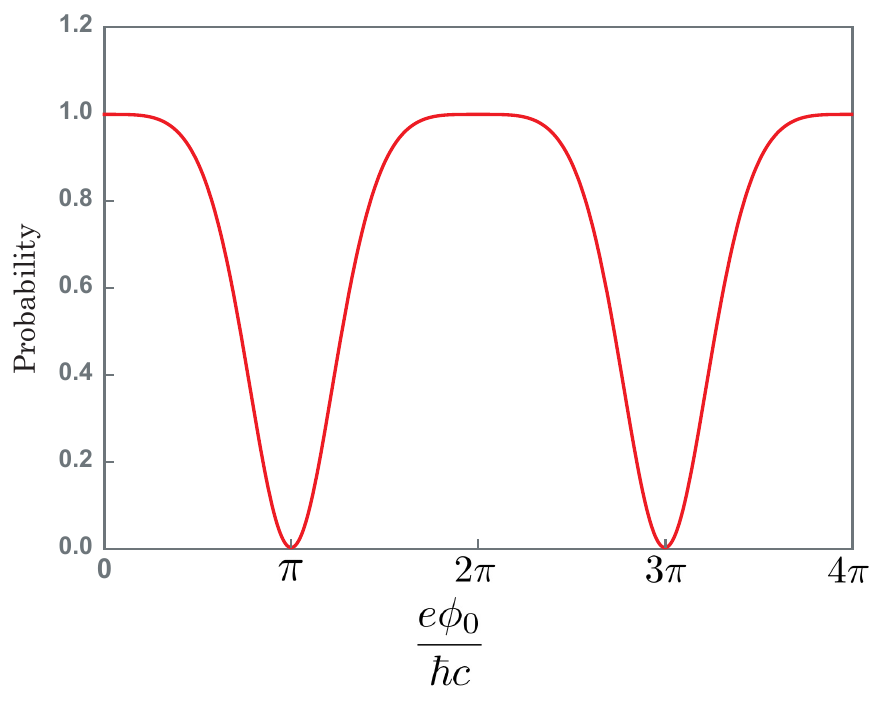}
  \caption{\label{fig6} Transmission probabilities vs magnetic flux when $k=\frac{\pi}{2}$.
   Here we adopt the same parameters in Fig.~\ref{fig5}. }
\end{figure}
\end{center}

For example, if we adopt the following parameters
\begin{equation}
 \alpha = \gamma = a=0, \quad \beta=\delta= b= \frac{\pi}{4} , \quad 
 L_{(1)}=L_{(2)}=0, \quad L_{(3)} \rightarrow \infty
\end{equation}
then Eqs.~(\ref{eq:me-R}) and (\ref{eq:me-T}) are reduced to
\begin{eqnarray}
 R & = & -\frac{e^{2ik \xi_{\rm I}}
   e^{2ik \left( \xi_{\rm I} - \xi_{\rm I\!I}  \right)} 
    \left( e^{-\frac{ie\phi_{0}}{\hbar c}} -1 \right)^2 }
   {e^{2ik \left( \xi_{\rm I} - \xi_{\rm I\!I}  \right)} \left( e^{-\frac{ie\phi_{0}}{\hbar c}} +1 \right)^2 
   - 4 e^{-\frac{ie\phi_{0}}{\hbar c}} }  ,  \\
 T & = &  
    \frac{2e^{2ik \left( \xi_{\rm I} - \xi_{\rm I\!I}  \right)}  e^{-\frac{ie\phi_{0}}{2\hbar c}}
    \left( e^{2ik \left( \xi_{\rm I} - \xi_{\rm I\!I}  \right)} -1 \right)
    \left( e^{-\frac{ie\phi_{0}}{\hbar c}} +1 \right) }
   {e^{2ik \left( \xi_{\rm I} - \xi_{\rm I\!I}  \right)} \left( e^{-\frac{ie\phi_{0}}{\hbar c}} +1 \right)^2 
    - 4 e^{-\frac{ie\phi_{0}}{\hbar c}} } .
\end{eqnarray}
In this case, while when $\displaystyle \frac{e\phi_{0}}{\hbar c}=2n\pi$ ($n=0,1,2, \cdots$),  we have $R=0$, 
when $\displaystyle \frac{e\phi_{0}}{\hbar c}=(2n+1)\pi$,  we have $T=0$
for every mode characterized by a wavenumber. 
In Fig.~\ref{fig5}, we show the transmission and reflection probabilities
as a function of the wavenumber $k$ in the cases of 
(a) $\frac{e\phi_{0}}{\hbar c}=0$, (b) $\frac{e\phi_{0}}{\hbar c}=\frac{\pi}{2}$,
   and (c) $\frac{e\phi_{0}}{\hbar c}=\pi$.
Furthermore,  Fig.~\ref{fig6} shows the transmission probability as a function 
of the magnetic flux when $k=\frac{\pi}{2}$.
Thus, the perfect transmission  and the perfect reflection appear
alternatively for every mode, as the strength of the magnetic flux increases.
Similar results may occur in a subclass in the parameter space in which $s_{11}=0$. 
(see Eqs.~(\ref{eq:sR}) and (\ref{eq:sT})).
Consequently, in this choice of parameters for the symmetric ring system, 
the modulation of the magnetic flux enable us to switch the current.


Finally, we comment on the previous work \cite{ald}.  In \cite{ald},
the authors discussed conductance, which is connected to transmission probability
via the Landauer formula, in a similar ring system composed of 
one-dimensional lattice using a tight-binding model.  
Although we have considered smooth one-dimensional wires, 
they considered the realistic lattice, whose system has
a dispersion relation different from that of a free particle.
However, when all the hopping integrals in \cite{ald} are taken to be 1 as in \cite{ald}, 
their system would correspond to a limited class of symmetric ring systems 
in which $s_{11}=s_{22}=s_{33}$ and $s_{12}=s_{21}=s_{13}=s_{31}=s_{23}=s_{32}$ in our model.
Thus our model deals with a very large class of junction conditions.

\section{Conclusion}
\label{sec:conclusion}

We have discussed quantum dynamics in
the quantum ring systems with double Y-junctions
in which two arms have same length. For these systems, 
a general formulation by scattering matrices
was found to be useful. Based on our formulations, 
we investigated localized states on the ring. 
Then we found that the symmetric ring systems 
in which one node has the same parameters as the other node 
under the reflectional symmetry possess the following interesting features.
{\it 
In the symmetric ring systems, localized states exist 
inevitably, and resonant perfect transmission 
occurs when the wavenumber of an incoming wave coincides 
with that of the localized states, for any parameters 
of the nodes except for the extremal cases in which 
the absolute values of components of scattering matrices take $1$.
}
We also investigated the external disturbance to the symmetric
ring systems. In particular,  we have considered the magnetic flux
penetrating the ring. 
Then we found that the current 
through the symmetric ring system for every mode characterized by a wavenumber 
can be switched simultaneously by the strength of the magnetic flux
only when we adopt a subclass of parameters for the Y-junctions.

We should briefly mention the time-reversal symmetry, 
which is a physically important class of problem, 
in the symmetric ring systems.
The time-reversal symmetry of the ring systems requires 
that the matrix $S_{\rm R}$ 
given by Eqs.~(\ref{eq:sr11})-(\ref{eq:sr22})
be symmetric. In the symmetric rings, 
using Eqs.~(\ref{eq:si-sii-r4c})-(\ref{eq:si-sii-r4c2}), we can show that 
$(S_{\rm R})_{12}=(S_{\rm R})_{21}$ from straightforward calculations.
Thus the time-reversal symmetry always holds 
in the symmetric ring systems.

In this paper, we did not deal with anti-symmetric rings,
which were considered in the previous work \cite{fknt}, 
and other cases in detail. 
Once we take a configuration which slightly deviates from 
the symmetric ring, numerous variations in the curves 
for transmission probabilities occur
due to the vast parameter space for the Y-junctions.
These would be investigated in the future works.
More general discussion on magnetic field effects
would also be provided.


\end{document}